\documentclass[runningheads]{llncs}
\usepackage{graphicx}
\usepackage{amsmath}

\DeclareMathOperator*{\argmin}{arg\,min}

\usepackage[margin=30mm, top=20mm, bottom=20mm]{geometry}

\begin{document}
\title{In-stream Probabilistic Cardinality Estimation for Bloom Filters}
%
%
\author{Rémy Scholler\inst{1,2} \and
Jean-François Couchot\inst{2} \and
Oumaïma Alaoui-Ismaïli\inst{1} \and
Denis Renaud\inst{3} \and
Eric Ballot\inst{4}}
\authorrunning{R. Scholler et al.}
%
\institute{Orange Innovation Orange Labs, Châtillon \and
Femto-ST Institute, DISC Department, UMR 6174 CNRS University of Bourgogne, Franche-Comté, Besançon \and
Orange Applications For Business Orange Labs, Belfort \and Mines ParisTech, CGS—Centre de Gestion Scientifique, Paris}
\maketitle              
\begin{abstract}
The amount of data coming from different sources such as IoT-sensors, social networks, cellular networks, has increased exponentially during the last few years. Probabilistic Data Structures (PDS) are efficient alternatives to deterministic data structures suitable for large data processing and streaming applications. They are mainly used for approximate membership queries, frequency count, cardinality estimation and similarity research. Finding the number of distinct elements in a large dataset or in streaming data is an active research area. In this work, we show that usual methods based on Bloom filters for this kind of cardinality estimation are relatively accurate on average but have a high variance. Therefore, reducing this variance is interesting to obtain accurate statistics. We propose a probabilistic approach to estimate more accurately the cardinality of a Bloom filter based on its parameters, i.e., number of hash functions $k$, size $m$, and a counter $s$ which is incremented whenever an element is not in the filter (i.e., when the result of the membership query for this element is negative). The value of the counter can never be larger than the exact cardinality due to the Bloom filter's nature, but hash collisions can cause it to underestimate it. This creates a counting error that we estimate accurately, in-stream, along with its standard deviation. We also discuss a way to optimize the parameters of a Bloom filter based on its counting error. We evaluate our approach with synthetic data created from an analysis of a real mobility dataset provided by a mobile network operator in the form of displacement matrices computed from mobile phone records. The approach proposed here performs at least as well on average and has a much lower variance (about 6 to 7 times less) than state of the art methods.

\keywords{Cardinality Estimation \and Bloom filters \and Streaming Data \and Probabilistic Data Structures \and Privacy}
\end{abstract}
\section{Introduction}
From the last few years, we are in the era of in-stream data \cite{StreamDataProba} and Internet of Things (IoT) \cite{IoT}. The amount of data coming from different sources such as IoT-sensors, social networks, cellular networks, has increased exponentially. The size (often in exabytes) and complexity of the data as well as the amount of noise associated with it is not predefined, making it hard to capture, store and process within the stipulated time. These type of data sets are usually known as Big data.

In \cite{singh_probabilistic_2020}, a definition of Big data's most relevant characteristics is proposed, and a focus is made on the use of Probabilistic Data Structures (PDS) for big data analytics. The efficient analysis of in-stream data often requires powerful tools such as Apache Spark~\footnote{https://spark.apache.org/} or Google Big Query~\footnote{https://cloud.google.com/bigquery?hl=en}. However, the use of deterministic data structures to perform analysis of in-stream data often include plenty of computational, space and time complexity. Probabilistic alternatives to deterministic data structures are better in terms of simplicity and constant factors involved in actual runtime. They are suitable for large data processing, approximate queries, fast retrieval and storing unstructured data, thus playing an important role in Big data processing, analysis and visualization \cite{singh_probabilistic_2020}. 

PDS are data structures having a probabilistic component \cite{PDSBD}. These probabilistic components are used to reduce time or space trade offs. PDS cannot give a definite answer, instead they provide with a reasonable approximation of the answer and a way to approximate this estimation. They are useful for Big data and streaming applications because they can decrease the amount of memory needed. In majority of the cases, these data structures use hash functions to randomize the items and ignore collisions to keep the size constant. That's one of the reasons why they cannot give exact values. According to \cite{singh_probabilistic_2020}, the main advantages of PDS are the small amount of memory they use (which can be controlled), their constant query time, the independence of hashes which makes them easily parallelizable and they are mainly used for approximate membership query, frequency count, cardinality estimation, and similarity research.

One of the most popular PDS are Bloom filters, that were proposed in \cite{BloomOriginal} as compact approximate set representations. The standard application of Bloom filters is for representing sets in a format suitable for answering membership queries, i.e., whether an element $x$ is member of a set $S$. Bloom filters enable answering membership queries in constant time, and with a configurable false positive probability. They improve upon alternative representations with respect to required memory and construction time. In particular, Bloom filter creation cost and memory requirements are linear with set size, with a very low constant. Bloom filters characteristics have been analyzed in depth, and several extensions have been proposed \cite{BloomVariants}. They can be used as cardinality estimators \cite{desfontaines_cardinality_2018}.

Finding the number of distinct values in a large dataset or in streaming data is an active research area because of its ever growing number of applications in a wide range of computer science domains \cite{harmouch_cardinality_2017}. Research has focused on reducing memory consumption and reading all data only once. Therefore, many algorithms aiming at approximating the cardinality of a dataset have been developed in this way \cite{harmouch_cardinality_2017}. Usual methods based on Bloom filters for cardinality estimation \cite{harmouch_cardinality_2017} are relatively accurate on average but have a high variance (this is experimentally shown in Section \ref{ExpeGlobal}). Therefore, reducing this variance is interesting to obtain accurate statistics.

We focus in this paper on the ability to estimate the number of distinct elements, from a large data stream, stored in a Bloom filter. There are several other data structures which address the same problem \cite{harmouch_cardinality_2017}. Compared to these works, our approach focuses on scenarios where a Bloom filter would anyway be required for membership testing, or is already available as it is the case in the context of one of our works on mining cellular network data to study human mobility while ensuring privacy. For these applications, our approach enables estimating the set cardinality with no additional cost.

In this work, we propose a probabilistic approach to estimate the exact cardinality of a Bloom filter based on its parameters, i.e., number of hash functions $k$, size $m$, and a counter $s$ which is incremented whenever an element is not in the filter (i.e., when the result of the membership query for this element is negative). The value of the counter can never be larger than the exact cardinality due to the Bloom filter's nature, but hash collisions can cause it to underestimate it. This creates a counting error that we estimate accurately, along with its standard deviation. The estimation is made in-stream and we consider the addition of elements from the stream one by one. We also discuss the way to optimize the parameters of a Bloom filter based on its counting error. We evaluate our approach with synthetic data created from an analysis of a real mobility dataset provided by a mobile network operator in the form of displacement matrices computed from mobile phone records. The approach proposed here performs at least as well on average, and has a much lower variance (about 6 to 7 times less).

The paper is structured as follows. In Section \ref{SotA} we review related work, and our probabilistic cardinality estimation approach is presented in Section \ref{ApproachGlobal}. Section \ref{ExpeGlobal} presents our experimentation and results and in Section \ref{Ccl} we conclude and discuss future work.

\section{Related Work}
\label{SotA}

In this Section we review related work through different aspects: Bloom filter basics (Section \ref{FBbasics}), and cardinality estimation with Bloom filters (Section \ref{CardEstimFB}). All the notations used in this Section and in Section \ref{ApproachGlobal} are summed up in the Table \ref{Notations}, which can be referred to if necessary.

\begin{table*}[ht]
  \centering
  \caption{Useful Notations and Definitions}
  \label{Notations}
  \begin{tabular}{ccl}
    \hline
    \hline
    Notation & Definition \\
    \hline
    \hline
    $|E|$ & Number of distinct elements in a universe $E$, i.e., cardinality of $E$ \\
    $m$ & Size of a Bloom filter (number of bits) \\
    $k$ & Number of hash functions of a Bloom filter \\
    $s$ & Incremental counter, defines also the filling state of a Bloom filter \\
    $BF(m, k, s)$ & Bloom filter with size $m$, number of hash functions $k$, in the filling state $s$ \\
    $B$ ($B_s$) & Number of bits set to $1$ (in the filling state $s$) \\
    $t_s$ & False positive probability of a Bloom filter filled with $s$ distinct elements \\
    $s_{max}$ & Number of distinct elements that can be added in a filter while keeping a false positive \\ &  probability below a target $t_{s_{max}}$  \\
    $\tilde{t_s}$ & Approximation of false positive probability of a Bloom filter filled with $s$ distinct elements \\
    $n$ ($n_s$) & Unique elements that should have been counted by the Bloom filter (in the filling state $s$) \\
    $\hat{n}$ ($\hat{n_s}$) & Literature estimators of $n$ and $n_s$ \\
    $U$ & Random variable equal to $1$ if a considered element has already been added to a filter \\
    & (true positive), and $0$ otherwise \\
    $p, p_s$ & Probability $P(U=0)$ \\
    $C$ & Random variable which gives the result of a filter check on a considered element \\
    $X_s$ & Counting error made by a Bloom  filter when moving from filling state $s$ to filling state $s+1$ \\
    $S_s$ & Counting error made by a Bloom filter in the filling state $s$, during its filling from $0$ to $s$ \\
    $N_s$ & Random variable which gives the corrected counter, $s + S_s$ \\
    \hline
    \hline
  \end{tabular}
\end{table*}

\subsection{Bloom Filter Basics}
\label{FBbasics}

Bloom filters are space-efficient PDS that were proposed in \cite{BloomOriginal} as compact approximate set representations. Considering a set $\{x_1, ..., x_n\}$ of $|E|$ distinct elements from a universe $E$, a Bloom filter consists of an array of $m$ bits and a family of $k$ independent hash functions $\{f_1, ..., f_k\}$, which associate to each element of $E$ an integer in the range of $[1,m]$.

\begin{figure}[htbp]
  \centering
  \includegraphics[width=0.6\columnwidth]{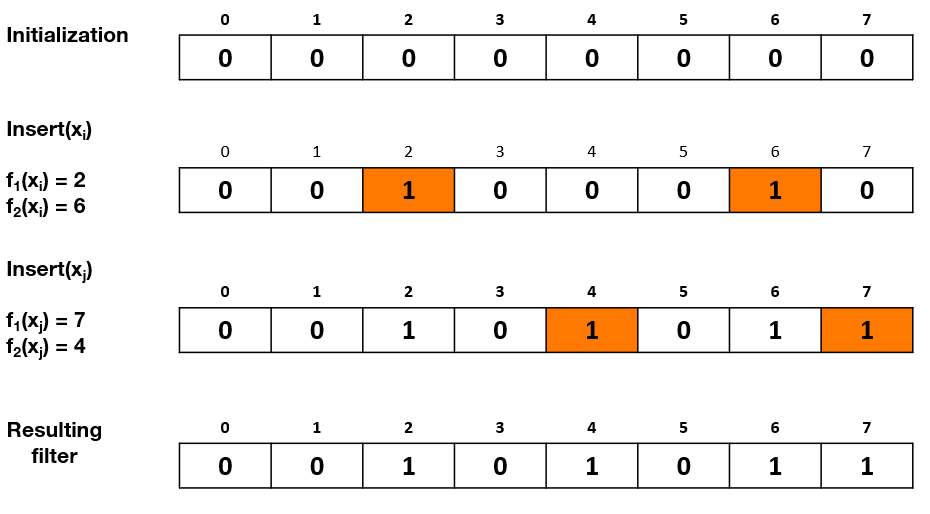}
  \caption{Initialization and insertions of elements in a Bloom filter with $m~=~8$ bits and $k~=~2$ hash functions.}
  \label{InsertionBloom}
\end{figure}

\begin{figure}[htbp]
  \centering
  \includegraphics[width=0.6\columnwidth]{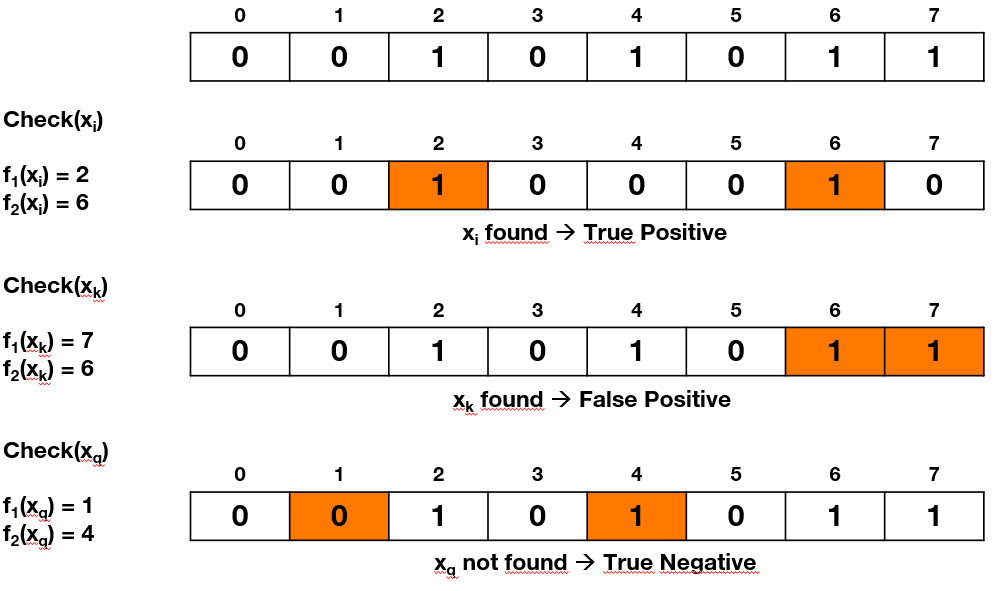}
  \caption{Checks of elements in a Bloom filter with $m~=~8$ bits and $k~=~2$ hash functions. The considered Bloom filter is the same as Figure \ref{InsertionBloom}.}
  \label{LookupBloom}
\end{figure}

All $m$ bits are set to $0$ (i.e., False) initially. An element $x$ is inserted into the Bloom filter by setting all positions $f_i(x)$, for $i \in {1, ..., k}$, of the bit array to $1$ (i.e., True). The Figure \ref{InsertionBloom} displays an initialization of a filter, and insertions of two distinct elements. We assume that an element $x$ is contained in the original set if all positions $f_i(x)$ of the Bloom filter are equal to $1$. If at least one of these positions is set to $0$, then we conclude that $x$ is not present in the original set. We say that an element $x$ is checked when we look at the values of the bits at the position $f_i(x)$. If all of these bits are set to $1$, the output is ``yes'', otherwise it is ``no''. One advantage of Bloom filters is that false negatives never occur. However, Bloom filters exhibit a small probability of false positives; due to hash collisions, it is possible that all bits representing a certain element have been set to $1$ by the insertion of other elements. The Figure \ref{LookupBloom} shows two checks of two distinct elements, with the resulting filter of Figure \ref{InsertionBloom}.

In \cite{FirstFPBF}, the false-positive probability is derived in the following way: The probability that any particular bit is equal to $0$ is $(1- \frac{1}{m})^{kn}$, since this value must be avoided by all $kn$ hash values. The probability that a particular bit is set to $1$ is then 

\begin{displaymath}
    t = 1 - (1- \frac{1}{m})^{kn}
\end{displaymath}

In order for an element to result in a false positive, each of its $k$ hash values must be the index of a bit that is set to $1$. The probability that this happens is then claimed to be, 
\begin{displaymath}
    t_n = t^k = (1 - (1- \frac{1}{m})^{kn})^k
\end{displaymath}
which can be approximated for large $m$ by,
\begin{equation}
    \tilde{t_n} \approx (1 - e^{-\frac{kn}{m}})^k
    \label{FPProb}
\end{equation}
thanks to $\ln(1+x)~\approx~x$ for $x~\to~0$.

However, it is assumed that event ``$ith$ bit is set to $1$'' and the event ``$1st$, $2nd$, ..., $(i-1)th$ bits are set to $1$'' are independent. This assumption is not always true, as it has been highlighted in \cite{bose2008false,christensen_new_2010}. Another analysis also arriving at approximation \eqref{FPProb} for the false positive probability, but without the previous assumption, can be found in \cite{Mitzenmacher_2005}.

The well-known formula \eqref{FPProb} is thus incorrect \cite{bose2008false,christensen_new_2010}. The exact false positive probability is derived in \cite{christensen_new_2010} which gives,

\begin{equation}
t_n = \frac{1}{m^{k(n+1)}} \sum_{i=1}^m i^k i! \binom{m}{i} 
\left\{ 
\begin{array}{l}
kn \\
    i 
\end{array}
\right\}
\label{ExactFPProb}
\end{equation}

where,

\begin{displaymath}
\left\{ 
\begin{array}{l}
kn \\
    i 
\end{array}
\right\} = \frac{1}{i!} \sum_{j=0}^i (-1)^{i-j} \binom{i}{j} j^{kn} 
\label{StirlingNumber}
\end{displaymath}
is the Stirling number of second kind~\footnote{en.wikipedia.org/wiki/Stirling\_numbers\_of\_the\_second\_kind}.
This expression is always larger than $\tilde{t_n}$ (defined in equation \eqref{FPProb}) and is numerically closed to $\tilde{t_n}$ for $m \gg k$, but not for smaller $m$ \cite{christensen_new_2010}. The exact false positive probability $t_n$ (defined in equation \eqref{ExactFPProb}) is not as simple to compute as $\tilde{t_n}$ but \cite{christensen_new_2010} derived a computationally solvable formula (even for large $m, n$ and $k$) based on a recursive expression. In the following, we consider the usual case $m \gg k$, so we keep up with the false positive probability $\tilde{t_n}$. However, in our approach, $\tilde{t_n}$ could be replaced by the exact false positive probability $t_n$ if necessary.

In some scenarios, the number $n$ of distinct elements stored in a Bloom filter is unknown. For these cases, the false positive probability is often based on the number of true bits (i.e., bits equal to $1$) in the filter. For a Bloom filter with $N$ bits set to true, the probability that one hash function points to a true bit equals to $\frac{N}{m}$. The probability that all $k$ hash functions point to true bits, which leads to a false positive, is then $(\frac{N}{m})^k$. That approximation is often used for cardinality estimation \cite{swamidass_mathematical_2007,papapetrou_cardinality_2010}.

The question of minimizing $t_n$ (defined in equation \eqref{ExactFPProb}) is usually not considered, since it is close to $\tilde{t_n}$ in common cases (i.e., $m \gg k$). For given number $n$ of distinct elements and a Bloom filter of length $m$, the false positive probability $t_n$ can be minimized by optimizing the ratio between true bits $B$ (i.e., bits equal to $1$) and Bloom filter length $m$. The false positive probability $\tilde{t_n}$ is minimized when this ratio is 0.5 \cite{papapetrou_cardinality_2010}. This is the case when the number of hash functions is set to,
\begin{equation}
    k \approx \frac{m}{n}\ln(2)
    \label{koptim}
\end{equation}  
For a given $n$, a target $t_n$, and assuming the optimal value for $k$, the number of bits $m$ can be computed with equation \eqref{FPProb}, leading to,
\begin{equation}
    m \approx -\frac{n\ln(t_n)}{\ln(2)^2}
    \label{mopt}
\end{equation} 
Therefore, with a given $n$ and a target $t_n$ it is possible to fix $m$ and $k$ together so they satisfy the two equations \eqref{koptim} and \eqref{mopt}. More in-depth analysis can be found in \cite{TargetFPProb,GoelGupta_2010}.

\subsection{Cardinality Estimation with Bloom Filters}
\label{CardEstimFB}

Here we focus on methods based on Bloom filters for cardinality estimation \cite{harmouch_cardinality_2017}. The standard Bloom filter is designed to maintain the membership information rather than a statistical information about the underlying dataset. But one can count the distinct elements in a multiset by combining a Bloom filter with a counter $s$, which is incremented whenever an element is not in the filter. The value of the counter can never be larger than the exact cardinality due to the Bloom filter's nature (i.e., hash collisions cause it to underestimate the number of unique elements). Bloom filters have been used effectively for cardinality estimation. The two main methods are those proposed in \cite{swamidass_mathematical_2007} and in \cite{papapetrou_cardinality_2010}. In \cite{swamidass_mathematical_2007}, an estimator $\hat{n}$ of the population of Bloom filter using $B$, the number of bits set to $1$ in the filter, has been introduced. Given a Bloom filter of the size $m$ with $k$ hash functions, the number of distinct elements can be estimated based on equation \eqref{FPProb} and an urn model such that,

\begin{equation}
    \hat{n} = -\frac{m}{k}\ln(1-\frac{B}{m})
    \label{UniqueSwamidass}
\end{equation}

In \cite{papapetrou_cardinality_2010}, the same type of estimator is proposed. The estimate is the maximum likelihood value for the number of hashed elements, based on \eqref{FPProb}. 

\begin{equation}
    \hat{n}  = \frac{\ln(1-\frac{B}{m})}{k\ln(1-\frac{1}{m})}
    \label{UniquePapapetrou}
\end{equation}
The authors prove that the Bloom filter configuration affects the estimation accuracy. Larger Bloom filter (i.e., large size $m$) provides higher estimation accuracy. Bloom filters with fewer hash functions exhibit a more accurate cardinality estimation.
Moreover, Bloom filters need a prior knowledge of the maximum cardinality in order to choose the suitable size of the filter. However, the main contribution of the cardinality estimation proposed in \cite{papapetrou_cardinality_2010} is the upper and lower bounds they derive from their estimator. One can notice that the previous two estimators are actually the same, for large $m$, at the first order in $\frac{1}{m}$, thanks to $\ln(1+x)~\approx~x$ for $x~\to~0$.

\section{Probabilistic Cardinality Estimation}
\label{ApproachGlobal}

In this Section, we detail our proposal for cardinality estimation with Bloom filters. We formulate our problem in Section \ref{ProbFor}, and propose an approach to solve it considering the addition of elements from the stream one by one in Section \ref{1par1Global}.

\subsection{Problem Statement}
\label{ProbFor}

State of the art approaches for cardinality estimation with Bloom filters \cite{harmouch_cardinality_2017} are accurate on average but have a high variance (see Section \ref{Exp1}). Our goal is thus to propose an accurate cardinality estimator with a lower variance.

\begin{figure}
  \centering
  \includegraphics[width=0.6\columnwidth]{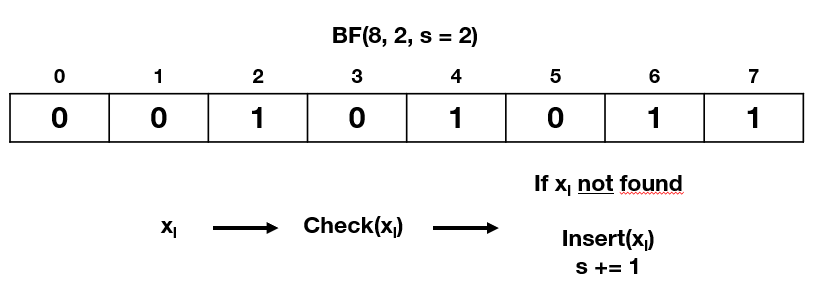}
  \caption{Possible insertion of an element in a Bloom filter with $m~=~8$ bits and $k~=~2$ hash functions combined with a counter $s$. The considered Bloom filter is the same as Figure \ref{InsertionBloom} and Figure \ref{LookupBloom}.}
  \label{New_Insertion_Bloom}
\end{figure}

In this work, we consider in-stream elements from a universe $E$ of size $|E|$ and we fill a Bloom filter from scratch. We note $t_s$ the false positive probability of a Bloom filter $BF(m, k, s)$, $m$ and $k$ being fixed when initializing the filter, and $s$ being a counter combined with the filter, which is incremented whenever the check of an element is negative. This counter thus corresponds to the number of elements that have modified the filter structure. We say that the filter $BF(m, k, s)$ is in the filling state $s$. When filling a filter, only the elements that change the structure of the filter (i.e., true negatives, which change at least one bit from $0$ to $1$) will be counted by $s$ and the others (i.e. true or false positives) will not. Each time the counter $s$ is incremented, the false positive probability $t_s$ increases. The filter thus ``forgets'' to count the false positives at each filling step. Therefore, the counter $s$ underestimate the true number of elements $n_s$ that should have been counted so far (i.e., when the filter reaches the filling state $s$). The Figure \ref{New_Insertion_Bloom} shows the possible insertion of an element $x_l$ in a Bloom filter with $m=8$ bits and $k~=~2$ hash functions combined with the counter $s$. The Bloom filter considered in this Figure is the same as in Figure \ref{InsertionBloom} and Figure \ref{LookupBloom}.

We thus propose an estimation of $n_s$, for each filling state $s$ ranging from $0$ to $s_{max}$. We note $s_{max}$ the maximum number of distinct elements that can be added in the filter while keeping a false positive probability below a targeted $t_{s_{max}}$. Obviously, we have $|E| \geq s_{max}$. The proposed estimator is based only on the false positive probability $t_s$ at the different filling states $s$ (on $\tilde{t_s}$ in practice, as explained in Section \ref{FBbasics}). We also propose an estimator for the standard deviation of the estimated number of false positives during the filter filling. Finally, this false positive counting added to the counter $s$ results in an accurate cardinality estimation, with associated error bounds, adapted to streaming data.

The parameters $m$ and $k$ are then set so $t_{s_{max}}$ is minimal ($s_{max}$ and $m$ being fixed). We choose this setup (see Section \ref{FBbasics}) because of the in-stream context. In a real situation, we may not be able to represent all the elements of a stream with only one Bloom filter because of its fixed size $m$. We could instead create another Bloom filter for with the same purpose when the performance of the first one decreases too much (i.e., its false positive probability is too high), as it is proposed in \cite{ScalableBF}.

In this work, we assume that the streaming elements are added one by one to the filter (see Section \ref{1par1Global}). One could also want to set a Bloom filter based on a cardinality estimation error (see Section \ref{OptSet}).

\subsection{Streaming elements added one by one}
\label{1par1Global}

In this part, we consider elements arriving one by one. We need to estimate the counting error made when moving from filling state $s$ to the next filling state $s+1$, i.e., estimate the number of false positives ``forgotten'' before another element is checked negative and added (thus modifying the filter structure). The random variable representing this error is noted $X_s$ and will be added to the other filling states errors. For all $s \in [1, s_{max}]$ we want to estimate, 
\begin{displaymath}
    S_s = \sum_{r=0}^{s-1} X_r
    \label{TotCountError}
\end{displaymath}
the random variable representing the counting error made by a filter in the filling state $s$, during its filling from $0$ to $s$. Obviously, $X_0 = 0$ and $S_1 = 0$ (thus $X_0$, $S_1$ are not random variables) because at the initial filling state, the Bloom filter is $BF(m, k, 0)$, i.e., empty. Whatever the first element, it is added to the filter with no error, and the Bloom filter becomes $BF(m, k, 1)$ (the associated false positive probability thus becomes $t_1$). Then, we note $N_s$ the random variable representing the corrected counter $s$, i.e., the true number of distinct elements that should have been counted so far (i.e., when the filter reaches the filling state $s$). Therefore, we have,
\begin{displaymath}
    N_s = s + S_s
    \label{CounterEstim}
\end{displaymath}

For each random element from the universe $E$ in the considered stream, we note $U$ the random variable which is equal to $1$ if the element has already been added to the filter (true positive), and $0$ otherwise. We note $C$ the random variable which gives the result of a check on this element, i.e., $0$ if the filter returns that the element has not been added previously (true negative) and $1$ otherwise (true positive, or false positive). A filling state change (i.e., $s$ is increased by $1$) occurs when an element raises a check $C$ equal to $0$. We note $p_s = P(U = 0)$ the probability that a random element has not been added to the filter $BF(m, k, s)$ previously. This ``prior'' probability depends mainly on $N_s$, the number of elements that should have been added to the filter, and on the universe $E$ (more precisely its cardinality $|E|$).

In what follows we consider two cases. In the first one we assume that there is no true positives, i.e., for all $s~\in~[1, s_{max}]$, $p_s~\approx~1$. This assumption could makes sense because the elements come from a potentially infinite universe (i.e., $|E| \to +\infty$, thus $\frac{|E|}{s_{max}} \gg 1$). In the second one, we consider that true positives can occur, and that each element from the universe $E$ has a uniform probability to be added to the filter. Therefore, we have for all $s~\in~[1, s_{max}]$, $p_s = 1 - \frac{N_s}{|E|}$. This probability thus decreases from $1$ (when the filter is empty, $s=0$) to $1 - \frac{N_{s_{max}}}{|E|}$ (when the filter reaches $s = s_{max}$).

\subsubsection{1st Case : Without True Positives}
\label{1par1.1}

Elements come from a potentially infinite universe, so we could consider that the chances of an item being added to the filter previously are very small, i.e., for all $s~\in~[ 1, s_{max} ]$, $p_s~\approx~1$.

With this assumption, if an element raises a value of $C$ equal to $1$, it is necessarily a false positive and for $s \in [ 1, s_{max} ]$, $X_s$ corresponds to the number of elements raising $C = 1$ \textit{before the first element raising $C = 0$} (i.e., the first element that modifies the filter's structure). We note $T_s$ the random variable which gives the number of elements necessary to obtain $C = 0$, and thus, 
\begin{displaymath}
    X_s = T_s - 1
\end{displaymath}

$T_s$ clearly follows a geometric distribution with parameter $1-t_s$ (which can be interpreted as the waiting time of the first $C=0$). Therefore we have, 

\begin{displaymath}
    \forall j \in \mathbf{N}^*, P(T_s = j) = t_s^{j-1}(1-t_s)
    \label{NoTP}
\end{displaymath}
leading to,
\begin{displaymath}
\begin{split}
    \mathbf{E}(T_s) & = \frac{1}{1-t_s} \\
    V(T_s) & = \frac{t_s}{(1-t_s)^2}
\end{split}
\end{displaymath}

Finally, the mean counting error for the filling state change $s$ to $s+1$ is,

\begin{displaymath}
    \mathbf{E}(X_s) = \mathbf{E}(T_s - 1) = \mathbf{E}(T_s) - 1 = \frac{t_s}{1-t_s}
    \label{NoTPE}
\end{displaymath}
and the associated variance is,
\begin{displaymath}
    V(X_s) = V(T_s - 1) = V(T_s) = \frac{t_s}{(1-t_s)^2}
    \label{NoTPV}
\end{displaymath}

We can notice that for $t_s << 1$, $\mathbf{E}(X_s)~\approx~t_s$ and $V(X_s)~\approx~t_s$ at the first order in $t_s$ (for small $t_s$).

Assuming mutual independence between the $X_s$ variables, we can deduce for all $s \in [ 1, s_{max} ]$,

\begin{equation}
\begin{split}
    \mathbf{E}(S_s) & = \sum_{r=1}^{s-1} \frac{t_r}{1-t_r} \\
    V(S_s) & = \sum_{r=1}^{s-1} \frac{t_r}{(1-t_r)^2}
\end{split}
    \label{NoTPTotalError}
\end{equation}
the mean counting error made by a filter in the filling state $s$, during its filling from $0$ to $s$, and its associated variance.

Therefore, we have, 

\begin{equation}
\begin{split}
    \mathbf{E}(N_s) & = s + \mathbf{E}(S_s) \\
    V(N_s) & = V(S_s) \\
\end{split}
    \label{CorrectedCounter}
\end{equation}
which are our estimation of $n_s$ and the variance associated.

\subsubsection{2nd Case: With True Positives}
\label{1par1.2}

We need to take in account for true positives because repetitions may occur in the stream. Therefore $1 - \frac{N_{s_{max}}}{|E|} < p_s < 1$, and for each element (as well as between different elements), for all $j \in \mathbf{N}^*$,
\begin{displaymath}
\begin{split}
    P(T_s = j) & = p_s(1-t_s)[1 - p_s + p_st_s]^{j-1} \\
    & = p_s(1-t_s)[1 - p_s(1-t_s)]^{j-1}
\end{split}
    \label{WTP}
\end{displaymath}
As in the previous case, $T_s$ (defined in Section \ref{1par1.1}) still follows a geometric distribution, but with parameter $1-p_s(1-t_s)$. However, in this case $X_s \ne T_s-1$ otherwise we would count true positives as counting errors.

Therefore, we can use the law of total probability,
\begin{displaymath}
    \forall r \in \mathbf{N}, P(X_s = r) = \sum_{j=1}^{+\infty}P(X_s=r, T_s=j)
    \label{WTPX}
\end{displaymath}
and, for all $r \in \mathbf{N}$, for all $j \in \mathbf{N}^*$, if $r \leq j-1$,
\begin{displaymath}
P(X_s = r, T_s = j) = \binom{j-1}{r}(p_st_{s})^{r}(1-p_s)^{j-1-r}p_s(1-t_s)
\label{WTPXcapT}
\end{displaymath} otherwise $P(X_s = r, T_s = j) = 0$.

These two last equations holds because $P(X_s = r, T_s = j)$ is the probability to have $r$ false positives and obtain the first true negative after $j$ elements (thus have obtained $j-1-r$ true positives).

We can then obtain, for all $r \in \mathbf{N}$,
\begin{displaymath}
\begin{split}
    P(X_s = r) & = \sum_{j=r+1}^{+\infty}\binom{j-1}{r}(p_st_{s})^{r}(1-p_s)^{j-1-r}p_s(1-t_s) \\
    & = (1-t_s)t_s^{r}\sum_{j=r+1}^{+\infty}\binom{j-1}{r}p_s^{r+1}(1-p_s)^{j-(r+1)} \\
\end{split}
\end{displaymath}
and recognizing on the right the sum of terms of a Pascal distribution with parameters $r+1$ and $p_s$, we can deduce

\begin{displaymath}
\forall r \in \mathbf{N}, P(X_s = r) = (1-t_s)t_s^r
\label{WTPFinal}
\end{displaymath}

The mean counting error for the filling state change $s$ to $s+1$ is thus,

\begin{displaymath}
    \mathbf{E}(X_s) = \frac{t_s}{1-t_s}
    \label{WTPE}
\end{displaymath}
and the associated variance is,
\begin{displaymath}
    V(X_s) = \frac{t_s}{(1-t_s)^2}
    \label{WTPV}
\end{displaymath}
thanks to the expectancy and variance of geometric distributions.

We therefore obtain the same results in mean and variance as in the first case. This result was to be expected because the true positives do not change the counting error but only lengthen the ``time" necessary to change the filter's filling state. That said, we can notice that we could also estimate the average number of true positives during a state change either by doing a similar calculation, or by counting the average number of elements needed to change state and subtracting $\mathbf{E}(X_s) + 1$.

Assuming mutual independence between the $X_s$ variables, we can deduce for all $s \in [ 1, n_{max} ]$,

\begin{equation}
\begin{split}
    \mathbf{E}(S_s) & = \sum_{r=1}^{s-1} \frac{t_r}{1-t_r} \\
    V(S_s) & = \sum_{r=1}^{s-1} \frac{t_r}{(1-t_r)^2}
\end{split}
    \label{ 2}
\end{equation}
the mean counting error made by a filter in the filling state $s$, during its filling from $0$ to $s$, and its associated variance.

Therefore, we have as in Section \ref{1par1.1}, 

\begin{equation}
\begin{split}
    \mathbf{E}(N_s) & = s + \mathbf{E}(S_s) \\
    V(N_s) & = V(S_s) \\
\end{split}
    \label{CorrectedCounter2}
\end{equation}
which are our estimation of $n_s$ and the variance associated.

\subsubsection{Optimal Setting}
\label{OptSet}

\begin{figure}
  \centering
  \includegraphics[width=0.6\columnwidth]{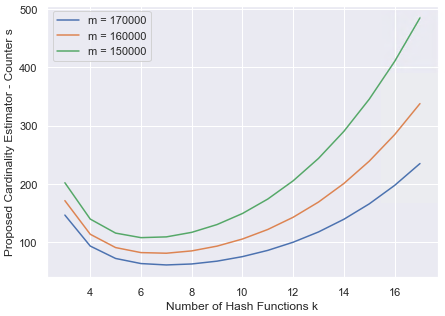}
  \caption{$\mathbf{E}(N_s) - s$ as a function of $k$. A minimum value exists for each $m$, which means an optimal value of $k$ can be found. For $m~=~160 000$ bits, this value is $k~=~6$ hash functions.}
  \label{OptHashFig}
\end{figure}

As explained in Section \ref{FBbasics}, the number of hash functions $k$ is usually fixed to minimize the false positive probability of the considered Bloom filter in the filling state $s_{max}$ (i.e., we do not expect to add more elements in the filter). It is also possible to target a false positive probability $t_{s_{max}}$ for a fixed $s_{max}$, and compute $m$, $k$ according to the targeted probability ensuring that $k$ is the optimal number of hash functions. Therefore, we remark that for $s_{max}$ and $m$ fixed, it is possible to compute $k_{opt}$ which minimize the mean counting error made by the Bloom filter in its filling from $s=0$ to $s = s_{max}$, considering that the filter is filled as in Section \ref{1par1Global}, i.e., element by element. Therefore, we look for,
\begin{displaymath}
\begin{split}
    k_{opt} & = \argmin\limits_{k \in [1,m]} (\mathbf{E}(N_s) - s) \\
    & = \argmin\limits_{k \in [1,m]} \sum_{s=0}^{s_{max}} \frac{t_s}{1-t_s}
    \label{kopt}
\end{split}
\end{displaymath}
For a given $s_{max}$, a target $\mathbf{E}(S_{s_{max}})$, and assuming $k~=~k_{opt}$, the number of bits $m$ (size) could be computed, at least numerically. In Figure \ref{OptHashFig}, we give an example showing the existence of $k_{opt}$ in this case, for fixed $s_{max} = 17 000$ and various values of $m$.

We notice as well that if we choose $k$ to minimize $t_{s_{max}}$ as usual, then $\mathbf{E}(S_{s_{max}})$ is bounded. An obvious upper bound is,
\begin{displaymath}
    \mathbf{E}(S_{s_{max}}) = \sum_{s=0}^{s_{max}} \frac{t_s}{1-t_s} \leq s_{max}\frac{t_{s_{max}}}{1-t_{s_{max}}}
    \label{UpperBound}
\end{displaymath}
because for fixed $m$ and $k$, the false positive probability $t_s$ is increasing with $s$ and $t_0~=~0$.

However, computing numerically $k$ and $m$ each time a Bloom filter is initialized may be not efficient enough for an in-stream context. In what follows, we did not choose this method to set Bloom filters, but we kept a setting described in Section \ref{FBbasics}.

\section{Experimentation}
\label{ExpeGlobal}

In this Section, we evaluate our approach with synthetic data created from an analysis of a real mobility dataset provided by a mobile network operator in the form of displacement matrices computed from mobile phone records. This dataset has been mined to study how mobility in France changed before and during lockdown in 2020 due to COVID-19 pandemic \cite{FLV}. These matrices comprised origin-destination travel flows among 1436 geographical areas of mainland France, stratified by day. Each area belongs to one of 13 regions, which are the subnational administrative divisions of mainland France. Mobile phone data were previously anonymized in compliance with strict privacy requirements, presented to and audited by the French data protection authority (Commission Nationale de l’Informatique et des Libertés).

We describe our experimental protocol in Section \ref{Protocol} and present our results in Section \ref{Exp1}. At the end of Section \ref{Exp1}, we discuss the limits of our model, for higher values of $s$ (i.e., $s \geq s_{max}$) allowing to reach higher values of $t_s$ (up to $10\%$). Finally, in Section \ref{Exp2} we present a variant of our proposal aiming at reducing the privacy leakage risk which is especially present at the beginning of the filter's filling, when the filter is ``almost empty''.

\subsection{Experimental Protocol}
\label{Protocol}

In \cite{FLV}, it is highlighted that after the lockdown of 17 March 2020, long-range mobility (i.e., $> 100$ km) almost completely stopped during weekends (94\% decrease, from 3 million to $0.17$ million trips per day, for daytime and night-time combined). We consider here that these $0.17$ million trips were made by cellular network user from the mobile phone dataset (though we do not have any information about these potential users). The idea is to store a fraction of these synthetic users in a Bloom filter considering them as a stream. The number $s_{max}$ is thus fixed to $10\%$ of $0.17$ million, i.e., $s_{max} = 17 000$. 

Therefore, we add elements in Bloom filters one by one in Section \ref{Exp1} until $s = s_{max}$ for each filter, and evaluate the approach explained in Section \ref{1par1Global}. We compare our cardinality estimator with the estimator \eqref{UniquePapapetrou} of Section \ref{CardEstimFB}. This means that for any filling state $s$, we consider the baseline estimator $\hat{n_s} = \frac{\ln(1-\frac{B_s}{m})}{k\ln(1-\frac{1}{m})}$, $B_s$ being the number of true bits of a Bloom filter in filling state $s$. The estimator defined by equation \eqref{UniqueSwamidass} being an approximation of $\hat{n_s}$ for large $m$ (see Section \ref{CardEstimFB}), we only compared our approach to $\hat{n_s}$. In the following, $\hat{n_s}$ is called the baseline estimator. To achieve this comparison, we used mainly three metrics, i.e., Mean Biased Error (MBE), Mean Absolute Error (MAE), and Root Mean Squared Error (RMSE). For the MBE and MAE, we also considered the associated standard deviations.

To account for various realistic situations, we consider different variation domains of the probability $p_s$ (see Section \ref{1par1Global}). We create different universes $E$ with fixed size $|E|$ depending on $s_{max}$, and sample elements with replacement so,
\begin{itemize}
    \item $|E| = \frac{1}{0.8}s_{max}$ such that $p_s$ decreases from $1$ to $p_{s_{max}} =  1 - \frac{s_{max}}{|E|} = 0.2$,
    \item $|E| = \frac{1}{0.6}s_{max}$ such that $p_s$ decreases from $1$ to $p_{s_{max}} = 0.4$,
    \item $|E| = \frac{1}{0.4}s_{max}$ such that $p_s$ decreases from $1$ to $p_{s_{max}} = 0.6$,
    \item $|E| = \frac{1}{0.2}s_{max}$ such that $p_s$ decreases from $1$ to $p_{s_{max}} = 0.8$.
\end{itemize}
To consider the case $p_s~\approx~1$ for all $s~\in~[1, s_{max}]$, we sample elements without replacement in a universe $E$ such that $|E| \gg s_{max}$ (to be sure to reach $s = s_{max}$). The case $p_{s_{max}}~\approx 0$ is not realistic because it would mean that there is only true positives after the filter's reaches $s = s_{max}$, i.e., the entire universe $E$ has been added to the filter.

For each variation domain of $p_s$, we consider 1000 Bloom filters with the same setting, i.e., $s_{max} = 17000$, $t_{s_{max}} = 0.01$ and the parameters $m$ and $k$ chosen as in Section \ref{FBbasics}. Then, 1000 universes (1 per filter) set according to the variation domain of $p_s$ (as explained above) are added to the filters, element by element (i.e., as a stream).

\subsection{Streaming Elements Added One By One}
\label{Exp1}

This Section shows the results of the proposition made in Section \ref{1par1Global}. The setting of the Bloom filters is $s_{max} = 17000$ and $t_{s_{max}} = 0.01$ which leads to $m = 162945$ and $k = 6$ according to Section \ref{FBbasics}. We considered five values of $p_{s_{max}}$, i.e., $p_{s_{max}} \in \{1, 0.8, 0.6, 0.4, 0.2\}$.

\begin{figure}[ht]
  \centering
  \includegraphics[width=0.8\columnwidth]{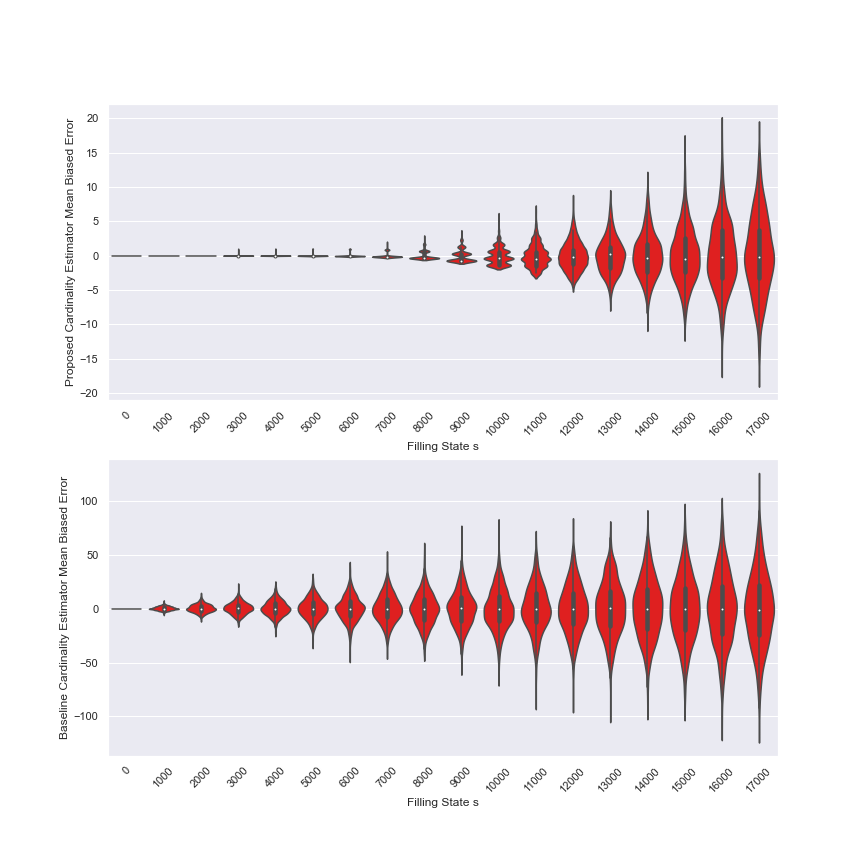}
  \caption{Biased error distribution of the baseline estimator (below) and our proposal (above), for $p_{s_{max}}=1$, as functions of $s$.}
  \label{p=1Violin}
\end{figure}

\begin{figure}[ht]
  \centering
  \includegraphics[width=0.5\columnwidth]{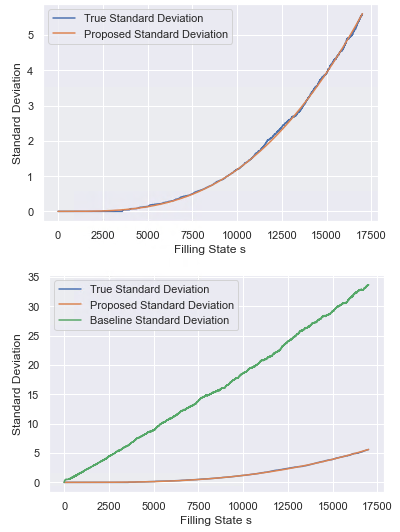}
  \caption{Standard deviation of the baseline estimator, our proposal, along with the true standard deviation of $n_s$, as functions of $s$, for $p_{s_{max}}=0.6$. Above, we removed the baseline standard deviation.}
  \label{Std_p=0.55}
\end{figure}

The Figure \ref{p=1Violin} corresponds to $p_{s_{max}}=1$. It shows the biased error of $\mathbf{E}(N_s)$, i.e., $\bar{n_s} - \mathbf{E}(N_s)$ (above), and the biased error of the baseline estimator \eqref{UniquePapapetrou} (below), as functions of $s$. Here, $\bar{n_s}$ is the mean over the 1000 Bloom filters of the true number of distinct elements $n_s$ seen by the 1000 Bloom filters when they are in filling state $s$. Introducing $\bar{n_s}$ (and the standard deviation of $n_s$) might be surprising but we recall that, for a filling state $s$, all of the 1000 Bloom filters does not have tried to count the same number of elements (i.e., for each $s$, there is one value of $n_s$ per filter). We also represent the distribution of the biased error as violin plots around the MBE, for each estimator (baseline and our proposal). We can see that the baseline estimator has much higher variance (about one order of magnitude) than our proposal. One advantage of this is that the proposed estimator's biased error is distributed very close to $0$ (with a standard deviation inferior to $1$) for a larger range of $s$ values than the baseline. This is also shown by the Figure \ref{Std_p=0.55}. The performance of the two approaches in terms of MBE are almost equivalent, our proposal only has a slightly lower MBE (for any $s$).

We can also see that the biased error distribution of the proposed estimator (Figure \ref{p=1Violin} above) is (almost) gaussian only for large enough values of $s$ (on the Figure \ref{p=1Violin}, it occurs at $s \approx 13 000$). We strongly believe this is due to the nature of $S_s = \sum_{r=0}^{s-1} X_r$, which is an sum of independant random variables $X_r$. These random variables are not identically distributed, but we believe that the Lindeberg (or Lyapunov) condition is satisfied. Therefore, we can apply the Lindeberg (or Lyapunov) Central Limit Theorem\footnote{https://en.wikipedia.org/wiki/Central\_limit\_theorem} (CLT), leading to this gaussian behaviour for large $s$. However, we intend to verify this in future work. We could also argue that the random variables $X_r$ are almost identically distributed since in our experimentation, for $s \in [0, 17000]$, $t_s \in [0, 0.01]$ which could be an interval small enough to argue that $\mathbf{E}(X_1) \approx ... \approx \mathbf{E}(X_s)$ so we can apply the classic CLT.

\begin{figure}[ht]
  \centering
  \includegraphics[width=0.8\columnwidth]{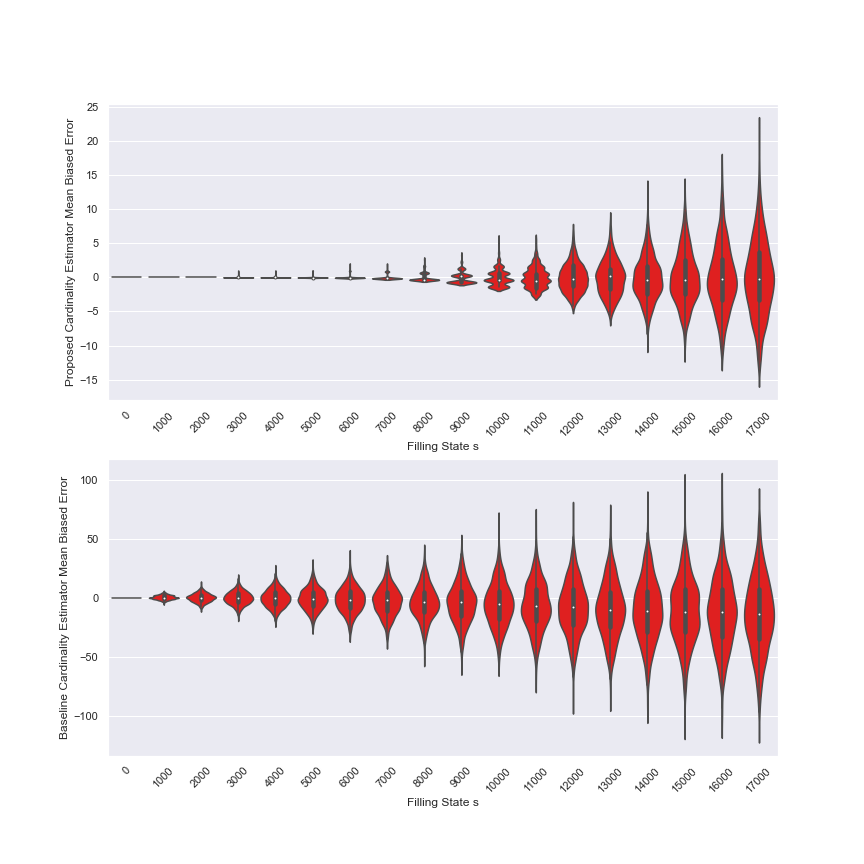}
  \caption{Biased error distribution of the baseline estimator (below) and our proposal (above), for $p_{s_{max}}=0.6$, as functions of $s$.}
  \label{p=0.55Violin}
\end{figure}

The Figure \ref{p=0.55Violin} is the same as \ref{p=1Violin} but for $p_{s_{max}}=0.6$, i.e., there is $45\%$ of duplicate elements in each one of the 1000 streams. The results and remarks are almost identical as the ones in Figure \ref{p=1Violin}, as expected from Section \ref{1par1.2}. Actually, the main difference between the $p_{s_{max}}=1$ and $p_{s_{max}}=0.6$ situations is the number of elements that are checked by the filters to reach $s = s_{max} = 17000$. For $p_{s_{max}}=1$, this number is $17032$ (average over the 1000 filters) whereas for $p_{s_{max}}=0.6$ it is $27287$.

\begin{figure}[ht]
  \centering
  \includegraphics[width=0.5\columnwidth]{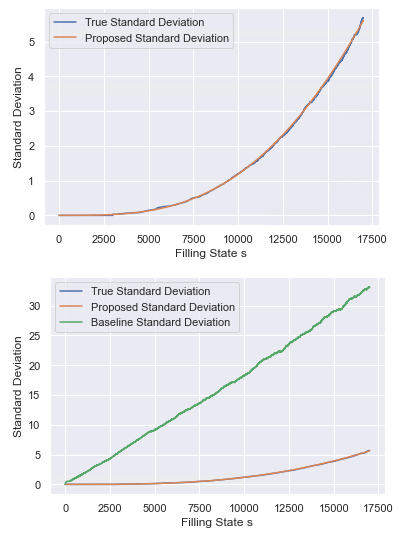}
  \caption{Standard deviation of the baseline estimator, our proposal, along with the true standard deviation of $n_s$, as functions of $s$, for $p_{s_{max}}=1$. Above, we removed the baseline standard deviation.}
  \label{Std_p=1}
\end{figure}

The Figure \ref{Std_p=0.55} shows the standard deviation of our proposal and baseline as functions of the filling state $s$, for $p_{s_{max}}=0.6$. We also represent the true standard deviation, i.e. the standard deviation of $n_s$, the true number of distinct elements seen by the 1000 Bloom filters when they are in filling state $s$. The baseline standard deviation is removed above on this figure. As in Figure \ref{p=1Violin} and \ref{p=0.55Violin}, we see that the baseline has a much higher standard deviation than our proposal. Moreover, the Figure \ref{Std_p=0.55} shows that our proposal for the standard deviation of $n_s$, i.e., $\sqrt{V(N_s)}$ (see Section \ref{1par1Global}), fits well with the true standard deviation of $n_s$. As noticed with Figures \ref{p=1Violin} and \ref{p=0.55Violin}, the standard deviation of our approach stays below $1$ for a larger range of $s$ values than the baseline (e.g., here it reaches almost $s=9000$ before exceeding $1$). The same analysis holds for $p_{s_{max}}=1$, as it is showed by Figure \ref{Std_p=1}.

The Tables \ref{TableMeanPerf} and \ref{TableMeanPerf2}, shows the MBE, MAE, RMSE and associated standard deviation (for MBE and MAE) of the baseline and our proposal, for $p_{s_{max}} \in \{1, 0.8, 0.6, 0.4, 0.2\}$, and $s = s_{max} = 17000$. We can see that for this filling state, corresponding to a false positive probability $t_s = 0.01$, our proposal is 6 to 7 times more accurate and less spread out than the literature baseline.

\begin{table*}[ht]
\centering
  \caption{Mean performances and associated standard deviation for $s=17000$, and $p_{s_{max}} \in \{1, 0.8, 0.6\}$}
  \label{TableMeanPerf}
  \begin{tabular}{ccc|cc|cc}
    \hline
      & \multicolumn{2}{c}{$p_{s_{max}}=1$} & \multicolumn{2}{c}{$p_{s_{max}}=0.8$} & \multicolumn{2}{c}{$p_{s_{max}}=0.6$}\\
    \hline
     & Baseline & Our Proposal
     & Baseline & Our Proposal
     & Baseline & Our Proposal \\
    \hline
    MBE & $-0.215 \pm 34.3$ & \textbf{0.21} $\pm$ \textbf{5.59} &
    $-3.28 \pm 33.5$ & \textbf{0.47} $\pm$ \textbf{5.68} &
    $-14.4 \pm 30.9$ & \textbf{-0.02} $\pm$ \textbf{5.52} \\
    \hline
    MAE & $27.1 \pm 21.0$ & \textbf{4.45} $\pm$ \textbf{3.38} &
    $26.5 \pm 20.7$ & \textbf{4.54} $\pm$ \textbf{3.44} &
    $27.3 \pm 20.5$ & \textbf{4.37} $\pm$ \textbf{3.38} \\
    \hline
    RMSE & $34.3$ & \textbf{5.59} &
    $33.6$ & \textbf{5.70} &
    $34.1$ & \textbf{5.52} \\
    \hline
  \end{tabular}
\end{table*}

\begin{table*}[ht]
\centering
  \caption{Mean performances and associated standard deviation for $s=17000$, and $p_{s_{max}} \in \{0.4, 0.2\}$}
  \label{TableMeanPerf2}
  \begin{tabular}{ccc|cc}
    \hline
     & \multicolumn{2}{c}{$p_{s_{max}}=0.4$} & \multicolumn{2}{c}{$p_{s_{max}}=0.2$} \\
    \hline
     & Baseline & Our Proposal
     & Baseline & Our Proposal\\
    \hline
    MBE & $-8.03 \pm 26.9$ & \textbf{0.08} $\pm$ \textbf{5.40} &
    $18.4 \pm 21.0$ & \textbf{1.24} $\pm$ \textbf{5.50} \\
    \hline
    MAE & $22.3 \pm 17.0$ & \textbf{4.27} $\pm$ \textbf{3.29} &
    $22.8 \pm 16.0$ & \textbf{4.49} $\pm$ \textbf{3.40} \\
    \hline
    RMSE & $28.0$ & \textbf{5.39} &
    $27.9$ & \textbf{5.63} \\
    \hline
  \end{tabular}
\end{table*}

\begin{figure}[ht]
  \centering
  \includegraphics[width=0.5\columnwidth]{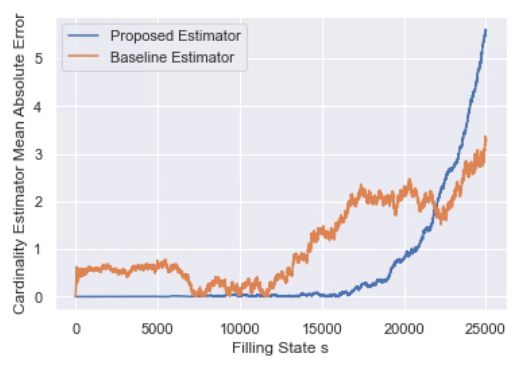}
  \caption{Mean Absolute Error (MAE) of our proposal and baseline as functions of $s$. This Figure shows the limits of our proposal in terms of mean perfomance, for large $s$ (i.e., large $t_s$).}
  \label{MAE_tvarie}
\end{figure}

Then we analyse the limits of our proposal. To do this, we continue to add elements to the Bloom filters so $s$ can exceed $s_{max} = 17000$, and thus $t_s$ can exceed $t_{s_{max}} = 0.01$. The Figure \ref{MAE_tvarie} shows the MAE of our proposal and baseline as functions of $s$. We can observe that until $s \approx 22000$, our proposal has a lower MAE than the baseline. The filling state $s \approx 22000$ corresponds to $t_s \approx 3\%$, which is relatively high in the context of this evaluation. The Figure \ref{MAE_tvarie} shows that for large values of $t_s$ (i.e., large values of $s$), the proposed cardinality estimator (above) is biased whereas the baseline (below) is not. Such high values of $t_s$ (greater than $1\%$) can easily be avoided by creating other filters with the same purpose (i.e., storing and counting the other elements of the stream).

\begin{figure}[ht]
  \centering
  \includegraphics[width=0.5\columnwidth]{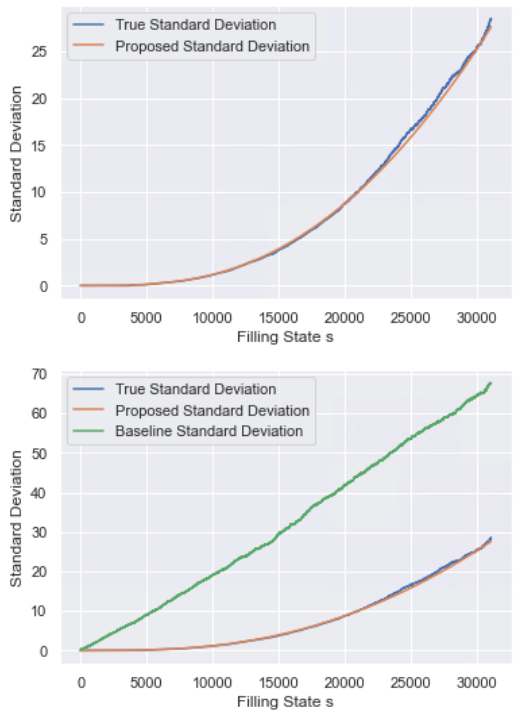}
  \caption{Standard deviation of the baseline estimator, our proposal, along with the true standard deviation of $n_s$, as functions of $s$, for $p_{s_{max}}=1$. Here, $s$ can exceed $s_{max}$, thus $t_s$ can exceed $t_{s_{max}}$. Above, the baseline standard deviation is removed.}
  \label{Std_t_varie}
\end{figure}

However, the Figure \ref{Std_t_varie} shows that our proposal has still a lower standard deviation than the baseline. It is important to notice that for $s \approx 30 000$, $t_s \geq 10\%$, which is a extremely high false positive probability in many applications. Moreover, it shows that our proposal for the standard deviation of $n_s$, i.e., $\sqrt{V(N_s)}$ (see Section \ref{1par1Global}), fits well with the true standard deviation of $n_s$, even for large values of $s$.

Finally, this experimentation shows that our proposal for cardinality estimation is accurate, and has much lower standard deviation than usual methods (6 to 7 times less according to Tables \ref{TableMeanPerf}) and \ref{TableMeanPerf2}. For common values of false positive probability (below a few percents), it is even more accurate than state of the art methods (about 5 times more accurate according to Tables \ref{TableMeanPerf} and \ref{TableMeanPerf2}). It has though its limits when reaching high values of false positive probability (above a few percents in the proposed experimentation), and the proposed estimator slightly biased. However, this bias is relatively low (i.e., about $0.001$ of relative MAE for $t_s = 10\%$, corresponding to $s=30000$ in our experimentation), and the proposed variance is still closer to reality than state of the art methods, and much lower than these methods. Moreover, high values of false positive probability (above 1\%) are usually avoided (it could be done by creating other Bloom filters for storing other elements of the same stream).

\subsection{First Step Towards Private Cardinality Estimation}
\label{Exp2}

We evaluated our approach with synthetic data created from an analysis of a real mobility dataset provided by a mobile network operator in the form of displacement matrices computed from mobile phone records. Mobile phone data were previously anonymized in compliance with strict privacy requirements, presented to and audited by the French data protection authority (Commission Nationale de l’Informatique et des Libertés). Anonymization is necessary in this usecase because location data is highly personal~\cite{UniqueCrowd} and its use comes with privacy issues~\cite{TrajMicroDataPriv}.

In some cases, cardinality estimation has to be made in a privacy preserving way because the raw data used to do the estimation cannot be easily anonymized. The raw data is then deleted after the cardinality estimation is done. We propose here a variant of our approach that aims at giving tools to handle compatibility with many legal frameworks, specifically with most of the interpretations of the European ePrivacy Directive, that controls the use of location data generated by mobile networks, combined with the classic European General Data Protection Regulation (GDPR). Whatever the allowance given by each regulator on its territory, one of the main driver is privacy leakage risk when storing location data history for each mobile user. This privacy leakage risk, is of course increasing with duration of stored data. To reduce this risk, we aim to avoid having a too low number of elements stored in a Bloom filter.

This especially happens at the beginning of the filter's filling. Therefore we propose the following filling method to mitigate the privacy leakage risk :
\begin{itemize}
    \item When the filter is still empty, store raw unique elements for a duration as long as possible, until the number of elements is fairly large. This number of elements depends on the usecase, and the duration discussed here has to be short and may be part of negotiations with legal authorities on each territory. Then, add them to the filter all at once. Because the filter is empty before the addition, it makes no false positives, and the counter $s$ does not need to be corrected.
    
    \item After this first step, continue the filter's filling by adding elements one by one as described before.
\end{itemize}
This leads to a lower mean counting error made by the filter during its filling from 0 to $s$, because it now has the form $E(S_s) = 0 + \sum_{r=b}^{s-1} \frac{t_r}{1-t_r}$, with $b$ the number of unique elements added during the first step (see Section \ref{1par1.2}).

We could then push further this method by adding elements by batches instead of adding them one by one after the first batch has been added. We did not work on this filling method because even if the theoretical mean counting error is intuitively lower with this method than with a one by one filling, the gain in terms of counting error is very low. To illustrate this, we tried a filling of $100$ filters until $s = 16970$ with a fixed batch size $b = 50$ elements. The experimental mean error made by the counters $s$ of the filters is then equal to $29.63$ whereas this same error with a one by one filling is $31.26$. The difference of mean counting error between the two methods at $s = 16970$ is less than $2$ elements, which is negligible at this filling state ($s = 16970$), and thus results in a negligible gain in bits per element. We noticed as well that the gain in counting precision is increasing with $b$ (batch size). However, high values of $b$ are not realistic because it is necessary to store raw personal data for a long time.

Finally, the method proposed here (i.e., a batch adding when the filter is still empty, and the adding elements one by one) for in-stream cardinality estimation is at least as precise as the method proposed in Section \ref{1par1Global}, and mitigates the risk of privacy leakage by allowing to avoid ``almost empty'' filters. The source codes of our proposals and experimentation will be publicly available online.

\section{Conclusion and Future Work}
\label{Ccl}

In this work, we showed that state of the art methods based on Bloom filters for the estimation of the number of distinct elements in a large dataset or in streaming data are relatively accurate in expectation but the variance is too high. We proposed a probabilistic approach to estimate the cardinality of a Bloom filter based on its parameters, i.e., number of hash functions $k$, size $m$, and a counter $s$ which is incremented whenever an element is not in the filter (i.e., when the result of the membership query for this element is negative). The estimation is made in-stream and we consider the addition of elements from the stream one by one. We also discussed the way to optimize the parameters of a Bloom filter based on its counting error. The whole approach is evaluated using a real mobility dataset provided by a mobile network operator in the form of origin-destination matrices computed from mobile phone records, aggregating the number of moves made from the origin to the destination. The approach proposed here performs at least as well on average, and has a much lower variance (about 6 to 7 times less) than common literature methods. This holds for a number of elements below a certain threshold corresponding to a false positive probability around $3\%$ (e.g., around 22000 elements for a filter with nearly 160000 bits and 6 hash functions).

We also designed a variant of our proposal to reduce the privacy leakage risk which is especially present at the beginning of the filter's filling, when the filter is ``almost empty''. As a side effect, this method also leads to a slightly lower cardinality estimation error.

In the context of our work, we will have to adapt our approach to privacy preserving data structures. It has been shown in \cite{desfontaines_cardinality_2018} that usual cardinality estimators do not preserve privacy. Bloom filters corresponds to the definition of cardinality estimators proposed in \cite{desfontaines_cardinality_2018}, so there is a need to modify this data structure in a privacy preserving way, as it has been done for example with flipped Bloom filters (BLIP) \cite{BLIP} and Pan-Private BLIP \cite{alaggan_privacy-preserving_2018} which are differentially private. In BLIP \cite{BLIP}, the idea is to flip the bits of a Bloom filter at random after it has been filled with real elements. In Pan-private BLIP \cite{alaggan_privacy-preserving_2018}, some bits are randomly set to $1$ before adding real elements, and when a real element is added, only a random fraction of the bits that should be set to $1$ are actually set to $1$. We do not intend to study the union and intersection of private Bloom filters because the cardinality estimation of a private intersection of Bloom filters has a extremely high variance (i.e., around 30\%) \cite{BloomCDR_OD}. Therefore, we intend to keep a high test membership utility and would be more interested in scalable Bloom filters \cite{ScalableBF} adapted to privacy constraints. The main idea in scalable Bloom filters is to fill another filter when the first one reaches a false positive probability threshold. Therefore, it could be interesting to work on private versions of these sorts of data structures and adapt our method to allow an accurate, private and scalable cardinality estimation.

\section*{Acknowledgment}

This work is a part of a research project carried out at Orange Innovation in collaboration with the Internet Physics Chair (Mines Paris - PSL University). This work is (partially) supported by the EIPHI Graduate School (contract ANR-17-EURE-0002).

\bibliographystyle{abbrv}
\bibliography{main}

\end{document}